\documentclass[aps,nofootinbib,notitlepage,superscriptaddress,twocolumn,10pt,prl]{revtex4-1}

\usepackage{graphicx}
\usepackage{amsmath,amssymb}
\usepackage{hyperref}
\usepackage{braket}
\usepackage{subfigure}
\usepackage{float}
\usepackage[dvipsnames]{xcolor}
\usepackage{mathrsfs}
\usepackage{comment}
\usepackage{multirow}
\usepackage{soul}

\def\S{\mathcal{S}}

\begin{document}

\title{
Unifying gravitational waves and dark energy}

\author{Alice Garoffolo}
 \email{garoffolo@lorentz.leidenuniv.nl}
\affiliation{Lorentz Institute for Theoretical Physics, Leiden University}

\author{Omar Contigiani}
 \email{contigiani@lorentz.leidenuniv.nl}
\affiliation{Lorentz Institute for Theoretical Physics, Leiden University}
\affiliation{Leiden Observatory, Leiden University}
\affiliation{Canadian Institute for Theoretical Astrophysics, University of Toronto}

\begin{abstract}
We present a unifying treatment for metric and scalar perturbations across different energy regimes in scalar-tensor theories of gravity.
To do so, we introduce two connected symmetry-breaking patterns: one due to the acquisition of nontrivial vacuum expectation values by the fields and the other due to the distinction between background and perturbations that live on top of it.  
We show that the geometric optics approximation commonly used to enforce this separation is not self-consistent for high-frequency perturbations since gauge transformations mix different tensor and scalar sectors orders.
We derive the equations of motions for the perturbations and describe the behavior of the solutions in the low and high-frequency limits.
We conclude by describing this phenomenology in the context of two screening mechanisms, chameleon and symmetron, and show that scalar waves in every frequency range are screened, hence not detectable.

\end{abstract}

\maketitle

Since the discovery of the Universe's late-time accelerated expansion, an incredible effort has been dedicated to understand its origin.
The most straightforward extension to the cosmological constant solution consists of a scalar field, $\phi$, coupled to the metric, $g_{\mu\nu}$, and acting 
as an additional gravitational force. 
Consequently, in these scalar-tensor theories (ST) the growth of matter perturbations and the propagation of gravitational waves (GWs) are modified.
Detecting $\phi$ via the former is a key goal of the next galaxy surveys \cite{Amendola:2016saw,LSST:2008ijt,Dore:2014cca,SKA:2018ckk} while the observation of GWs by LIGO-Virgo (LVK) \cite{LIGOScientific:2016aoc} has opened the possibility of using the latter. 
%
The evolution of matter perturbations on the large scales, where linear perturbation theory holds, is traditionally studied using of an effective field theory (EFT) covering all ST theories with second-order equations of motion (EoMs) \cite{Bloomfield:2012ff,Gubitosi:2012hu,Gleyzes:2013ooa}.
The EFT assumes a Friedmann-Robertson-Walker (FRW) background field configuration and regards the scalar field as the Goldstone boson of the time translation' spontaneous symmetry breaking (SSB). It prescribes the most general, low-energy action for the metric perturbations, both scalar and tensor. This is accomplished in the unitary gauge, where the scalar field perturbations are set to zero and are eaten by the metric. 
The presence of $\phi$ changes the dynamics of the gravitational potentials, possibly introducing scale dependencies in the growth of cosmic structures, and the EFT formalism is able to produce predictions for the modified observables \cite{Hu:2013twa, Raveri:2014cka, Zumalacarregui:2016pph, Kobayashi:2019hrl}.
%
On the other hand, GWs in General Relativity and ST theories are usually addressed in the high energy limit \cite{Isaacson:1967zz,Isaacson:1968zza,Garoffolo:2019mna,Dalang:2020eaj}. 
By introducing an expansion in derivatives of high-frequency (HF) perturbations, it is possible to study the propagation of GWs over slow-varying, but otherwise unknown, backgrounds.
The separation of variation scales between perturbation and background can be regarded as another SSB and helps to identify the true degrees of freedom (dofs) of the theory 
\cite{Maggiore:1900zz,Bluhm:2016fjc}. For example,
ST theories predict the presence of an additional scalar wave (SW) \cite{Hou:2017bqj,Eardley:1974nw,Zumalacarregui:2016pph} and possibly introduce extra damping \cite{LISACosmologyWorkingGroup:2019mwx, Belgacem:2018lbp,Garoffolo:2019mna,Tasinato:2021wol} or modifications to the propagation speed of the modes \cite{Mirshekari:2011yq}. 
Moreover, the HF expansion is well suited to describe the GWs observed by LVK and third-generation interferometers \cite{Punturo:2010zz,Reitze:2019iox} because their frequency can be as high as the EFT energy cutoff, and their sources may be located close enough for the spacetime geometry not to be FRW. 

Combining the information obtained via these two observables, GWs and matter perturbations, is a compelling task in light of the future scientific missions, although nontrivial due to the differences in energy scales and formalism used to describe them. 
For instance, the combined detection of GW170817 and GRB170817A \cite{LIGOScientific:2017vwq} set the speed of GWs to the one of light. However, the frequency of this event was close to the EFT cutoff, so its implications on the EFT parameter space are unclear \cite{deRham:2018red}. 
Additionally, all viable ST theories must be equipped with a screening mechanism to suppress the force carried by the scalar field in high-density regions such as the Solar System, where all the tests performed exclude its presence \cite{Adelberger:2003zx, Will:2014kxa}. The role of screening in shaping the distribution of matter in the Universe and constrain these theories has been studied extensively \cite{Jain:2010ka, Koyama:2015vza, Baker:2019gxo}.
And while such mechanisms have been discussed \cite{Ezquiaga:2020dao, Dalang:2020eaj} in the context of the HF expansion, it is still not understood whether an SW would pass through a screened region and be detected on Earth. 

This paper presents a formalism able to interpolate between the EFT description in the low energy regime and the HF expansion in the high energy limit. We study the dynamics of metric and scalar perturbations in an ST theory over a background that spontaneously breaks one spacetime symmetry, as in the EFT approach, and rigorously introduce the HF expansion, interpreting it as a consequence of the SSB. 
The key feature of our approach
is the definition of perturbations themselves.
While the eft literature defines them as the difference between the entire fields and their fixed background configurations, we define the perturbations via their null vacuum expectation values (vevs), inspired by the averaging procedure discussed in \citep{Isaacson:1968zza}. 
Once developed, we aim to apply this formalism to two screening mechanisms, chameleon and symmetron, to understand whether the SW would be detectable.

\smallskip
{\it Spontaneous symmetry breaking.---}
We consider a subset of 
generalized Brans-Dicke theories \cite{Brans:1961sx,DeFelice:2010jn} for which the action in Einstein frame (EF) can be written as the canonical action for a scalar field, $\phi$, coupled to the matter fields, $\Psi_i$, through a conformal transformation: 
\begin{multline}\label{eq:action}
    \S = \int \text{d}^4 x\, \sqrt{- g}\, \left[R - \frac12 g^{\mu\nu}\,\nabla_\mu \phi \nabla_\nu \phi - V(\phi) \right] + \\ + \S_m ( \Omega^2(\phi) g_{\mu\nu}, \Psi_i).
\end{multline}
In this expression, $R$ is the Ricci scalar and $V(\phi), \, \Omega(\phi)$ are two arbitrary functions modeling the field potential and the conformal coupling. 
We study the dynamics of perturbations in vacuum. This assumption is justified, for example, in cosmological settings when the scalar field dominates the energy content of the Universe and drives the expansion of the Universe, or in screening scenarios outside a localized matter field. Generalizing this assumption will be the topic of future works.
We note that our results do not cover theories exhibiting the Vainshtein screening mechanism \cite{Vainshtein:1972sx, Babichev:2013usa}, since they cannot be cast in the form of Eq.~\eqref{eq:action}.

Breaking spontaneously the symmetries of the action~\eqref{eq:action}, the fields acquire non trivial vevs
\begin{equation}\label{eq:vevs}
    \braket{g_{\mu\nu}} = \bar{g}_{\mu\nu} \,, \quad \braket{\phi} = \bar \phi\,.
\end{equation}
We require $\bar \phi$ to be function of the spacetime coordinates in order to define the preferred vector field
\begin{equation}
     v_\mu \,\equiv \, \partial_\mu \bar \phi \neq 0\,
\end{equation}
and the order parameter $L$
\begin{equation}\label{eq:DefL}
    L \, \equiv \, \sqrt{|v^\mu v_\mu|}\,.
\end{equation}
We assume that $\{\bar g_{\mu\nu}, \bar \phi \}$ vary on the same length scale and, without loss of generality, we take $|\bar{g}_{\mu\nu}| \,, |\bar \phi| \sim {\cal O} (1)$, so that $L$ measures the variation length scale of $\bar \phi$. 

Introducing an orthonormal tetrad such that ${v^a = \,e^a_\mu \, v^\mu}$, under diffeomoprhisms and local Lorentz transformations $v^\mu$ transforms as
\begin{eqnarray}\label{eq:BrokenSym}
(v_\mu)' &=& \partial_\mu  \braket{\phi - \xi^\rho\, \partial_\rho \phi } = v_\mu -  \partial_\mu (v_\rho\braket{\xi^\rho})\,, \\
(v^a)' &=& \Lambda^a_b \,e^b_\mu \, \partial^\mu \braket{ \phi } =\Lambda^a_b \, v^b\,, \end{eqnarray}
where $\Lambda^a_b$ is a Lorentz matrix and $\xi_\mu$ the generator of spacetime translations. The broken symmetry transformations are those such that $\Lambda^a_b \, v^b \neq 0$ and $v_\rho \braket{\xi^\rho} \neq 0$.
Note that $v^\mu \neq 0$ is crucial to have an SSB since the case $\bar \phi = const$ contains the maximally symmetric Minkowski, de Sitter and anti-de Sitter solutions.

\smallskip
{\it Definition of field perturbations.---}
We study the dynamics of the field perturbations
around their vevs
\begin{equation}\label{eq:fieldexpansion}
g_{\mu\nu} = \bar g_{\mu\nu} + h_{\mu\nu} \,, \quad\, \phi = \bar \phi + \delta \varphi \,,
\end{equation}
defined via $\braket{h_{\mu\nu}} = \braket{\delta \varphi} = 0$, compatibly with Eq.~\eqref{eq:vevs}.
We also assume that the amplitude of the perturbations is smaller than their background counterparts. This is quantified by the parameter $\alpha$ such that ${|h_{\mu\nu}| \sim |\delta \varphi| \sim \alpha \ll 1}$.
To describe the behaviour of oscillatory perturbations, such as GWs and SWs, we also define 
\begin{equation} 
    \epsilon \equiv \frac{\lambda}{L}\,,
\end{equation}
where $\lambda$ is the order of magnitude of the derivatives of the perturbations: $|\partial h_{\mu\nu}|, |\partial \delta \varphi| \sim 1/ \lambda $.
This parameter allows us to introduce the {\it ADM averaging scheme} \cite{Isaacson:1968zza,Isaacson:1967zz} to formally evaluate the vevs $\braket{\dots}$: oscillatory perturbations average out to zero after integrating over volumes that are larger than $\lambda$ but small enough to be independent of $\epsilon$. 
In practice, $\epsilon$ is used to separate between the so-called {\it low-frequency modes}, i.e., the background, and the {\it high-frequency modes}, i.e., the oscillatory perturbations. We note that the very existence of $\{h_{\mu\nu}, \delta \varphi \}$ requires $\epsilon < 1$, otherwise they would become part of the background as the perturbation's wavelength $\lambda$ approaches the background's length-scale $L$. The limit $\epsilon \rightarrow 1$ is particularly subtle since the volumes that need to be considered to make the ADM averages might become too big. 
Because the amplitudes of $\{h_{\mu\nu}, \delta \varphi \}$ can be made large or small via a gauge transformation and $\{\bar g_{\mu\nu}, \bar \phi \}$ are unknown, the requirement $\alpha \ll 1$ is not enough to distinguish background from perturbation as in Eq.~\eqref{eq:fieldexpansion} \cite{ Maggiore:1900zz}. This is the principal reason for introducing the parameter $\epsilon$: perturbations and backgrounds are distinguished according to their different variation length scales via the averaging scheme used to take the vevs. 

To discuss the role of diffeomorphisms, we decompose the vector field generating the gauge transformations as
\begin{equation}
    \label{eq:Xiseparation}
    \xi^\mu = \bar \xi^\mu + \delta \xi^\mu\,,
\end{equation}
where $| \delta \xi^\mu | \lesssim \alpha$ and $|\partial\delta \xi^\mu|\sim 1/\lambda$, such that ${\braket{\delta \xi^\mu } = 0}$. This way $\bar \xi^\mu$ generates the gauge transformations of $\{\bar g_{\mu\nu}, \bar \phi \}$, while $\delta \xi^\mu$, those of the HF perturbations $\{h_{\mu\nu}, \delta \varphi \}$. The field perturbations transform as
\begin{eqnarray}
  h'_{\mu\nu} &=& h_{\mu\nu} - (\bar \nabla_\mu \delta \xi_\nu + \bar \nabla_\nu \delta \xi_\mu ) \label{eq:GaugeH},\\
  \delta \varphi' &=& \delta \varphi - v^\mu \delta \xi_\mu, \label{eq:GaugePhi}
\end{eqnarray}
where $\bar \nabla_\mu$ is the covariant derivative associated to $\bar g_{\mu\nu}$. 
To preserve the splitting of Eq.~\eqref{eq:fieldexpansion} we restrict the class of allowed HF diffeomorphisms requiring 
\begin{equation}\label{eq:Xicondition}
    |\bar \nabla_\mu \delta \xi_\nu|\, ,  \,  |v^\mu \delta  \xi_\mu| \, \lesssim \, \alpha\,,
\end{equation}
so that the amplitudes of the field perturbations after the gauge fixing are $ \lesssim \alpha$.
From 
Eq.~\eqref{eq:Xicondition} we then see that
\begin{equation}\label{eq:XiconditionEpsilon}
|\partial_\mu \delta \xi_\nu| \sim \frac{|\delta \xi_\nu|}{\lambda} \sim \frac{|\delta \xi_\nu|}{\epsilon} \lesssim \alpha \, \rightarrow \, |\delta \xi_\nu| \lesssim  \epsilon \, \alpha\,.
\end{equation}
This is what we mean by second symmetry breaking: depending on the value of $\epsilon$, not every HF diffeomorphism is allowed  \cite{Maggiore:1900zz,Bluhm:2016fjc}. This requirement
is not imposed in the EFT cosmological perturbation theory since the quantities $\{\bar g_{\mu\nu}, \bar \phi \}$ are assumed a priori and the perturbations are uniquely defined as $h_{\mu\nu} = g_{\mu\nu} - \bar g_{\mu\nu} $ or $\delta \varphi = \phi - \bar \phi$. Consequently, the EFT is able to describe perturbations varying on every length-scale, even those close to the background when $\epsilon \sim 1$, provided that they are below the energy cutoff.
Conversely the traditional HF treatment does not assume a background, but works only in the $\epsilon\ll 1$ regime.  Our vev-based definitions allow us to bridge the gap between the high-energy/HF and low-energy/EFT treatments by describing perturbations around unknown backgrounds in the entire $\epsilon < 1$ regime. This is why the formalism presented here acts as missing link between these two approaches.




{\it High-frequency expansion.---} We can use $\epsilon$ to set up the expansions\footnote{Since it can be shown that $\alpha \ll \epsilon$ \cite{Maggiore:1900zz}, we expand only to first order in $\alpha$.}
\begin{eqnarray}
 h_{\mu\nu } &= h^0_{\mu\nu} + \epsilon \,  b_{\mu\nu} \,, \quad b_{\mu\nu} &= h^{1}_{\mu\nu} + \epsilon h^{2}_{\mu\nu} + \dots, \label{eq:hExpansion}\\
 \delta \varphi &= \delta \varphi^0 + \epsilon \, \psi  \,, \quad \psi &= \delta \varphi^{1} + \epsilon \delta \varphi^{2}+ \dots,  \label{eq:PhiExpansion}\\
 \delta \xi_\mu &= \delta \xi^0_\mu + \epsilon \, \delta \zeta_\mu  \,, \quad   \delta \zeta_\mu  &=  \delta \xi^{1}_\mu  + \epsilon  \delta \xi^{2}_\mu + \dots, \label{eq:XiExpansion}
\end{eqnarray}
where $|h^i_{\mu\nu}| \sim |\delta \varphi^i| \sim |\delta \xi^i_\mu| \sim \alpha$. 
If the fields perturbations satisfy a wave equation,
one can assume the WKB ansatz where $\{h^0_{\mu\nu}, \delta \varphi^0 \}$ coincide with the geometric optics (GO) order terms and $\{b_{\mu\nu}, \psi \}$ with the beyond geometric optics corrections \citep{Isaacson:1967zz,Isaacson:1968zza}. 
When $\epsilon \ll 1$ Eqs.~\eqref{eq:hExpansion}-\eqref{eq:XiExpansion} are meaningful perturbative expansions and condition~\eqref{eq:XiconditionEpsilon} leads to
\begin{equation}\label{eq:XiZerocondition}
    \delta \xi^0_\mu = 0 \,.
\end{equation}
Moreover, when $\epsilon \ll 1$ the gauge transformations can be reorganized in powers of $\epsilon$,
\begin{eqnarray}
 h_{\mu\nu}' &=& h_{\mu\nu} - \epsilon (\bar \nabla_\mu \delta \zeta_\nu + \bar \nabla_\nu \delta \zeta_\mu) \,, \label{eq:gaugehleadingorders}\\
 (\delta \varphi^0)' &=& \delta \varphi^0 \,,\\
 \psi' &=& \psi - v^\mu \delta \zeta_\mu\,, \label{eq:gaugepsileadingorders}
\end{eqnarray}
from which we see that $\delta \varphi^0$ is gauge invariant and that the leading order terms transform as
\begin{equation}
  (h^0_{\mu\nu})' = (h^0_{\mu\nu}) - \epsilon (\partial_\mu \delta \xi^1_\nu + \partial_\nu \delta \xi^1_\mu) \,, \quad (\delta \varphi^0 )' = \delta \varphi^0\,,
 \end{equation}
i.e., as if they lived on a flat background. This is not surprising, since covariant derivatives commute when acting on perturbations approximated at leading order in $\epsilon$ \cite{Isaacson:1967zz}.

Eqs.~\eqref{eq:gaugehleadingorders} and \eqref{eq:gaugepsileadingorders} show that  diffeomorphisms mix $h_{\mu\nu}^0$ and the second-order $\delta \varphi^1$ whenever $v_\mu \neq 0$, i.e. in the presence of an SSB. 
Therefore, fixing $h_{\mu\nu}^0$ generates $\delta \varphi^1 \neq 0$, even if one started by neglecting it. Vice versa, keeping only the leading orders term of HF expansion may lead to inconsistencies because this implicitly assumes  $\psi = 0$, using up one of the gauge freedom and leaving one less to fix $h_{\mu\nu}$.  
We conclude that keeping only the leading orders of the HF expansion, namely $\{h_{\mu\nu}^0, \delta \varphi^0 \}$, is an inconsistent approximation scheme.

In contrast, $\delta \phi$ is gauged away at every order in the EFT formalism. In our framework, this is reproducible in the limit $\epsilon \lesssim 1$ where Eq.~\eqref{eq:XiconditionEpsilon} becomes trivial. 
The difference between these two behaviours, namely at $\epsilon \ll1$ versus $\epsilon \lesssim1$, proves that some gauges choices are not suitable to describe perturbations across different energy scales.
Known gauge-invariant quantities, such as the Bardeen's potentials \citep{Bardeen:1980kt},
fall in this category. 


\smallskip
{\it Gauge Fixing and Equations of motion.---}
Assuming ${v^\mu v_\mu > 0}$ we define the orthogonal projector\footnote{This choice is suitable to investigate screening in a spherically symmetric and static spacetime. In cosmological settings where $v^\mu v_\mu < 0$ then $\Lambda_{\mu\nu} \equiv \bar{g}_{\mu\nu} + n_\mu n_\nu$. } 
\begin{equation}\label{eq:DefProjector}
    \Lambda_{\mu\nu} \equiv \bar{g}_{\mu\nu} - n_\mu n_\nu \,, \quad n_\mu \equiv \frac{v_\mu}{L}\,,
\end{equation}
and decompose the metric perturbation as
\begin{equation}
     h_{\mu\nu} = n_\mu n_\nu A + (n_\mu B_\nu + n_\nu B_\mu) + C_{\mu\nu}\,,
\end{equation}
with $A \equiv n^\rho n^\sigma h_{\rho \sigma}$, $B_\mu \equiv n^\rho \Lambda^\sigma_\mu h_{\rho \sigma}$ and $C_{\mu\nu} \equiv \Lambda^\rho_\mu \Lambda^\sigma_\nu h_{\rho \sigma}$ \cite{Tasinato:2021wol}. 
We impose the conditions
\begin{equation}\label{eq:ADMgauge}
    A=0\,, \qquad  B_\mu = 0 .
\end{equation}
Then, using the residual gauge freedom, we also fix:
\begin{equation}\label{eq:Gaugegauge}
    C = 0 \,, \qquad \bar \nabla^\mu C_{\mu\nu} = 0\,.
\end{equation}
Note that we have exhausted the gauge freedom since Eqs.~\eqref{eq:ADMgauge} and ~\eqref{eq:Gaugegauge} amount to $4$ conditions each. $B_\mu$ only has $3$ independent components being orthogonal to $n^\mu$ and the condition $C=0$ implies $n^\mu \bar \nabla^\nu C_{\mu\nu}=0$, in fact
\begin{equation}
   n^\mu \bar \nabla^\nu C_{\mu\nu} = - C_{\mu\nu} (K^{\mu\nu} - n^\mu a^\nu) = - C_{\mu\nu} K^{\mu\nu} \propto C,
\end{equation}
where $K_{\mu\nu} \equiv \Lambda^\rho_\mu \bar \nabla_\rho n_\nu$. and $a_\mu \equiv n^\rho \bar \nabla_\rho n_\mu$.
In the last step, we used the fact that scalars can be computed in any coordinate system and that choosing $\bar{\phi}$ as a coordinate implies $K_{\mu\nu} \propto \Lambda_{\mu\nu}$.
Using the background equations of motion (EoMs),
\begin{eqnarray}\label{eq:BGeq}
  \bar R_{\mu\nu} = \frac{1}{2} \left(v_\mu v_\nu + \bar g_{\mu\nu} \bar V \right) \,, \qquad
  \bar \Box \bar \phi = \bar V' \,.
\end{eqnarray}
 where $\bar V = V(\bar \phi)$ and $\bar V' = (\partial V/ \partial \phi)|_{\bar \phi} $, one can show that the combination
\begin{equation}\label{eq:gaugeInvQuant}
    \big(\bar \Box \delta \varphi  - \delta \varphi  \,  \bar V'' \big) +  \bar \nabla^\mu \, [v^\nu h_{\mu\nu}] - \frac{1}{2} \, v^\mu \bar \nabla_\mu \, h
\end{equation}
is gauge invariant, where $\bar V'' = (\partial^2 V/ \partial^2 \phi)|_{\bar \phi} $, and $h$ is the trace of $h_{\mu\nu}$.  The last two quantities in the equation above vanish in the chosen gauge, hence $\delta \varphi$ is invariant under the residual gauge freedom and different from zero. In the HF limit this result concerns $\psi$ since $\delta \varphi^0$ is already invariant.

We expand the action~\eqref{eq:action} to second order in the perturbations and find the EoMs
\begin{align}
&\bar{\Box} \gamma_{\mu\nu} + 2 \bar R_{\lambda\mu\alpha\nu} \gamma^{\lambda \alpha} = 0 \label{eq:Gamma},\\
&\bar \Box \,\delta \varphi - \delta \varphi \,\bar V''= 0, \label{eq:Phi}  
\end{align}
where we renamed the metric perturbation after the gauge fixings $\gamma_{\mu\nu}$.
This system of equations is valid for every value of $\epsilon<1$ and represents a spin 2 wave, $\gamma_{\mu\nu}$, and a spin 0 wave, $\delta \varphi$.
Our gauge choice clarifies which are the dofs, especially highlighting that $\delta \varphi^0$ and $\psi$ are not independent.
Eqs~\eqref{eq:Gamma},~\eqref{eq:Phi} do not display the damping term typical of non-minimally coupled ST theories \cite{LISACosmologyWorkingGroup:2019mwx} because they describe perturbations in EF.
This factor can be recovered by going to Jordan Frame (JF) where matter is coupled to the JF metric $\tilde{g}_{\mu\nu} \equiv \Omega^{2} (\phi) \,{g}_{\mu\nu}$. Moreover, we do not find modifications in the propagation speed of the modes because this effect is not predicted in the ST theories considered here.

\smallskip
{\it High-frequency limit.---}
We study the  $\epsilon \ll 1$ limit of the EoMs above to study HF GWs and SWs. Following the considerations illustrated when discussing the gauge transformations, we keep the first non-null orders of the $\epsilon$ expansions~\eqref{eq:hExpansion}-\eqref{eq:PhiExpansion}.
Since all of the dofs satisfy a wave equation, we make the following WKB ansatz
\begin{equation}\label{eq:GOansatz}
    \gamma_{\mu\nu} = \Upsilon_{\mu\nu}\, e^{i \theta/ \epsilon}\,,  \quad \delta \phi^0 = \Phi  \, e^{i \theta/ \epsilon} \,, \quad \psi = \Psi\, e^{i \theta/ \epsilon}\,,
\end{equation}
where $\Upsilon_{\mu\nu}, \Phi, \Psi$ are complex and of order ${\cal O}(\epsilon^0)$, $\theta$ is real and they are all slow-varying functions of the spacetime coordinates.
Because a derivative acting on the exponential brings down a $1/\epsilon$ factor, we can separate the EoMs into their $\epsilon^{-2}$, $\epsilon^{-1}$ and $\epsilon^{0}$ orders.
The leading order gives
\begin{equation}\label{eq:geodesiceq}
   \bar g^{\mu\nu}\, k_\nu\, k_\mu = 0\,,
\end{equation}
where $k_\mu \equiv \partial_\mu \theta$ is the wave vector.
Therefore $k_\mu$ is a null vector tangent to a null geodesic $k^\mu \bar \nabla_\mu k_\nu = 0$ which are interpreted as the rays of the graviton and scalar bundles \cite{Isaacson:1967zz,Isaacson:1968zza}.
At orders $\epsilon^{-1}$ and $\epsilon^0$ we find
\begin{align}
    & 2 k^\rho \,\bar \nabla_\rho \Upsilon_{\mu\nu} + \Upsilon_{\mu\nu} \bar \nabla_\rho k^\rho = 0, \\
    & 2 k^\rho \,\bar \nabla_\rho \Phi + \Phi \,\bar \nabla_\rho k^\rho = 0,
    \label{eq:Phiamp}
    \\
    & 2 k^\rho \,\bar \nabla_\rho \Psi + \Psi \,\bar \nabla_\rho k^\rho = i (\bar \Box \Phi - \Phi \,\bar V'' ).
    \label{eq:Psiamp}
\end{align}
The equations above imply that the squared amplitudes of ($\Upsilon_{\mu \nu}, \Phi$) scale with the inverse cross sectional area of the particle's bundle, while $\Psi$ has an additional imaginary source/sink term.

\smallskip
{\it Observables.---}
We can understand the effect of the gravitational and scalar waves on test particles by looking at the geodesic deviation equation in JF.

In the $\epsilon \ll 1 $ limit, the JF metric perturbation is $\delta \tilde{g}_{\mu\nu} \equiv \tilde{H}_{\mu\nu} e^{i \theta/\epsilon}$ with
\begin{equation}\label{eq:HE}
    \tilde{H}_{\mu\nu} = \Omega^2(\bar \phi) \left[  \Upsilon_{\mu\nu} + 2 \bar g_{\mu\nu}\, \frac{ \Omega'({\bar \phi})}{ \Omega({\bar \phi})}\, (\Phi + \epsilon \Psi)\right]\,
\end{equation}
and the perturbation of the JF Riemann tensor is
\begin{align}
\label{eq:Rie}
    \delta &\tilde{R}_{\mu\nu\rho\sigma} = -  \frac{2}{\epsilon^2} \, k_{[\rho} k_{[\nu}\tilde{H}_{\mu]\sigma]} \,  e^{i \theta / \epsilon} + \nonumber \\
    & + \frac{2i}{\epsilon} \left[ k_{[\rho} \tilde{\nabla}_{[\nu} \tilde{H}_{\mu]\sigma]} +  k_{[\nu} \tilde{\nabla}_{[\rho} \tilde{H}_{\sigma]\mu]}  + \tilde{H}_{[\mu[\sigma} \tilde{\nabla}_{\rho]} k_{\nu]}\right]  e^{i \theta / \epsilon},
\end{align}
where the square brackets stand for antisymmetrization and $\tilde{\nabla}_{\mu}$ is the covariant derivative associated to the background JF metric. 
We have verified that the JF Riemann tensor perturbation is invariant under both JF and EF gauge transformations up to order $\epsilon^{-1}$, as it should since it is related to observables.
The acceleration between two nearby geodesics is given by the contraction of Eq.~\eqref{eq:Rie} with $\tilde{u}^\mu \tilde{u}^\rho$, the four-velocity of a JF observer.
Being $\Upsilon_{\mu\nu}$ orthogonal to $v_\mu$ and not $\tilde{u}^\mu$, it could be that the polarization content seen by the observer is different than the standard case. Investigating this possibility will be the topic of further works. 

\smallskip
Finally, we discuss the detectability of the SW in a screened region. In the low-frequency regime (${\epsilon^2 \gtrsim 1/{\bar V}''}$), it has been shown that Eq.~\eqref{eq:Phi} describes a damped wave \cite{Katsuragawa:2019uto}. Hence, waves in this energy range are screened. 
In the HF regime (${\epsilon \ll 1}$), one has to use Eqs.~\eqref{eq:Phiamp},~\eqref{eq:Psiamp} which show that $\Phi$ is not affected by the background configuration of the scalar field, implying that a HF SW can pass through a screened region. However, the interaction with observers is regulated by the geodesic deviation equation. From Eq.~\eqref{eq:HE} we see that the SW contribution to $\delta \tilde{R}_{\mu \nu \rho \sigma}$ is multiplied by $\Omega'(\bar \phi)/\Omega(\bar \phi)$, whose form depend on the type of screening mechanism.
We consider two cases: chameleon and symmetron. Inside screened regions, the former requires $\Omega'/\Omega \sim 1/M$ where $M \sim 10^{-5}$ in units of Planck mass \cite{Khoury:2003rn,Burrage:2017qrf}, while in the latter requires $\Omega'/\Omega \sim 0$ \citep{Hinterbichler:2011ca}. %
Hence, we conclude that SWs would not be detectable in the high-energy limit because their interaction with matter is suppressed.

\smallskip
{\it Conclusions.---}
The growth of matter perturbations and the propagation of GWs are two essential probes of the source of the late-time cosmic accelerated expansion, which must be used jointly. However, they span two distinct energy ranges, and the assumptions used to describe them are very different. 
The formalism introduced here, based on the parameter $\epsilon$, reproduces the results of the low-energy EFT regime (in the range ${\epsilon \lesssim 1}$), where $\delta \varphi$ can be entirely gauged away, and naturally includes the short wavelength limit (${\epsilon \ll 1}$) probed by GWs. We connected these two approaches to describe perturbations of an ST theory via two related symmetry-breaking patterns: the acquisition of nontrivial vevs by the fields and the separation of the high- and low-frequency modes.
Working in the EF, we showed that the ST theory \eqref{eq:action} exhibits three propagating dofs $\{\gamma_{\mu\nu}, \delta \varphi \}$ and
discussed how the commonly used first-order GO approximation is not self-consistent in the presence of an SSB. This is because the next to leading order scalar field perturbation $\psi$ mixes with the leading metric perturbation $h^0_{\mu\nu}$ via the gauge transformations. 
We then derived the general perturbed EoMs.~\eqref{eq:Gamma},~\eqref{eq:Phi} and applied them to discuss the detectability of the additional scalar dof through GW observations. We investigated the cases of chameleon and symmetron screenings, and concluded that the SWs present in these theories are not detectable on Earth no matter their wavelength. In the low-frequency limit, when ${\epsilon^2 \gg 1/V''}$, 
the SW is damped 
by the nontrivial background profile. While this is not true when $\epsilon \ll 1$, we showed that screening suppresses the interactions between the SW and matter via the multiplicative factor $\Omega'(\bar \phi)/ \Omega(\bar \phi)$ in Eq~\eqref{eq:Rie}, making the SW undetectable also in this case.
Therefore we conclude that a direct detection of a scalar wave inside a screened region would systematically rule out ST theories based on chamaleon or symmetron screenings. If scalar waves are not detected, one may still use other propagation effects to probe viable theories of gravity via GW observations \cite{LISACosmologyWorkingGroup:2019mwx,Belgacem:2018lbp,Garoffolo:2020vtd,Tasinato:2021wol, Zumalacarregui:2016pph}.

{\it Acknowledgements.---}
We thank Alessandra Silvestri and Gianmassimo Tasinato for useful discussion.
AG acknowledges support from the NWO and the Dutch Ministry of Education. OC is supported by an NWO de Sitter fellowship and the Natural Sciences and Engineering Research Council of Canada (NSERC). 
\bibliographystyle{unsrtnat}
\bibliography{references.bib}

\begin{thebibliography}{44}
\providecommand{\natexlab}[1]{#1}
\providecommand{\url}[1]{\texttt{#1}}
\expandafter\ifx\csname urlstyle\endcsname\relax
  \providecommand{\doi}[1]{doi: #1}\else
  \providecommand{\doi}{doi: \begingroup \urlstyle{rm}\Url}\fi

\bibitem[Amendola et~al.(2018)]{Amendola:2016saw}
Luca Amendola et~al.
\newblock {Cosmology and fundamental physics with the Euclid satellite}.
\newblock \emph{Living Rev. Rel.}, 21\penalty0 (1):\penalty0 2, 2018.
\newblock \doi{10.1007/s41114-017-0010-3}.

\bibitem[Ivezi\'c et~al.(2019)]{LSST:2008ijt}
\v{Z}eljko Ivezi\'c et~al.
\newblock {LSST: from Science Drivers to Reference Design and Anticipated Data
  Products}.
\newblock \emph{Astrophys. J.}, 873\penalty0 (2):\penalty0 111, 2019.
\newblock \doi{10.3847/1538-4357/ab042c}.

\bibitem[Dor\'e et~al.(2014)]{Dore:2014cca}
Olivier Dor\'e et~al.
\newblock {Cosmology with the SPHEREX All-Sky Spectral Survey}.
\newblock 12 2014.

\bibitem[Bacon et~al.(2020)]{SKA:2018ckk}
David~J. Bacon et~al.
\newblock {Cosmology with Phase 1 of the Square Kilometre Array: Red Book 2018:
  Technical specifications and performance forecasts}.
\newblock \emph{Publ. Astron. Soc. Austral.}, 37:\penalty0 e007, 2020.
\newblock \doi{10.1017/pasa.2019.51}.

\bibitem[Abbott et~al.(2016)]{LIGOScientific:2016aoc}
B.~P. Abbott et~al.
\newblock {Observation of Gravitational Waves from a Binary Black Hole Merger}.
\newblock \emph{Phys. Rev. Lett.}, 116\penalty0 (6):\penalty0 061102, 2016.
\newblock \doi{10.1103/PhysRevLett.116.061102}.

\bibitem[Bloomfield et~al.(2013)Bloomfield, Flanagan, Park, and
  Watson]{Bloomfield:2012ff}
Jolyon~K. Bloomfield, \'Eanna~\'E. Flanagan, Minjoon Park, and Scott Watson.
\newblock {Dark energy or modified gravity? An effective field theory
  approach}.
\newblock \emph{JCAP}, 08:\penalty0 010, 2013.
\newblock \doi{10.1088/1475-7516/2013/08/010}.

\bibitem[Gubitosi et~al.(2013)Gubitosi, Piazza, and Vernizzi]{Gubitosi:2012hu}
Giulia Gubitosi, Federico Piazza, and Filippo Vernizzi.
\newblock {The Effective Field Theory of Dark Energy}.
\newblock \emph{JCAP}, 02:\penalty0 032, 2013.
\newblock \doi{10.1088/1475-7516/2013/02/032}.

\bibitem[Gleyzes et~al.(2013)Gleyzes, Langlois, Piazza, and
  Vernizzi]{Gleyzes:2013ooa}
Jerome Gleyzes, David Langlois, Federico Piazza, and Filippo Vernizzi.
\newblock {Essential Building Blocks of Dark Energy}.
\newblock \emph{JCAP}, 08:\penalty0 025, 2013.
\newblock \doi{10.1088/1475-7516/2013/08/025}.

\bibitem[Hu et~al.(2014)Hu, Raveri, Frusciante, and Silvestri]{Hu:2013twa}
Bin Hu, Marco Raveri, Noemi Frusciante, and Alessandra Silvestri.
\newblock {Effective Field Theory of Cosmic Acceleration: an implementation in
  CAMB}.
\newblock \emph{Phys. Rev. D}, 89\penalty0 (10):\penalty0 103530, 2014.
\newblock \doi{10.1103/PhysRevD.89.103530}.

\bibitem[Raveri et~al.(2014)Raveri, Hu, Frusciante, and
  Silvestri]{Raveri:2014cka}
Marco Raveri, Bin Hu, Noemi Frusciante, and Alessandra Silvestri.
\newblock {Effective Field Theory of Cosmic Acceleration: constraining dark
  energy with CMB data}.
\newblock \emph{Phys. Rev. D}, 90\penalty0 (4):\penalty0 043513, 2014.
\newblock \doi{10.1103/PhysRevD.90.043513}.

\bibitem[Zumalac\'arregui et~al.(2017)Zumalac\'arregui, Bellini, Sawicki,
  Lesgourgues, and Ferreira]{Zumalacarregui:2016pph}
Miguel Zumalac\'arregui, Emilio Bellini, Ignacy Sawicki, Julien Lesgourgues,
  and Pedro~G. Ferreira.
\newblock {hi\_class: Horndeski in the Cosmic Linear Anisotropy Solving
  System}.
\newblock \emph{JCAP}, 08:\penalty0 019, 2017.
\newblock \doi{10.1088/1475-7516/2017/08/019}.

\bibitem[Kobayashi(2019)]{Kobayashi:2019hrl}
Tsutomu Kobayashi.
\newblock {Horndeski theory and beyond: a review}.
\newblock \emph{Rept. Prog. Phys.}, 82\penalty0 (8):\penalty0 086901, 2019.
\newblock \doi{10.1088/1361-6633/ab2429}.

\bibitem[Isaacson(1968{\natexlab{a}})]{Isaacson:1967zz}
Richard~A. Isaacson.
\newblock {Gravitational Radiation in the Limit of High Frequency. I. The
  Linear Approximation and Geometrical Optics}.
\newblock \emph{Phys. Rev.}, 166:\penalty0 1263--1271, 1968{\natexlab{a}}.
\newblock \doi{10.1103/PhysRev.166.1263}.

\bibitem[Isaacson(1968{\natexlab{b}})]{Isaacson:1968zza}
Richard~A. Isaacson.
\newblock {Gravitational Radiation in the Limit of High Frequency. II.
  Nonlinear Terms and the Ef fective Stress Tensor}.
\newblock \emph{Phys. Rev.}, 166:\penalty0 1272--1279, 1968{\natexlab{b}}.
\newblock \doi{10.1103/PhysRev.166.1272}.

\bibitem[Garoffolo et~al.(2020)Garoffolo, Tasinato, Carbone, Bertacca, and
  Matarrese]{Garoffolo:2019mna}
Alice Garoffolo, Gianmassimo Tasinato, Carmelita Carbone, Daniele Bertacca, and
  Sabino Matarrese.
\newblock {Gravitational waves and geometrical optics in scalar-tensor
  theories}.
\newblock \emph{JCAP}, 11:\penalty0 040, 2020.
\newblock \doi{10.1088/1475-7516/2020/11/040}.

\bibitem[Dalang et~al.(2021)Dalang, Fleury, and Lombriser]{Dalang:2020eaj}
Charles Dalang, Pierre Fleury, and Lucas Lombriser.
\newblock {Scalar and tensor gravitational waves}.
\newblock \emph{Phys. Rev. D}, 103\penalty0 (6):\penalty0 064075, 2021.
\newblock \doi{10.1103/PhysRevD.103.064075}.

\bibitem[Maggiore(2007)]{Maggiore:1900zz}
Michele Maggiore.
\newblock \emph{{Gravitational Waves. Vol. 1: Theory and Experiments}}.
\newblock Oxford Master Series in Physics. Oxford University Press, 2007.
\newblock ISBN 978-0-19-857074-5, 978-0-19-852074-0.

\bibitem[Bluhm(2016)]{Bluhm:2016fjc}
Robert Bluhm.
\newblock {Gravity with background fields and diffeomorphism breaking}.
\newblock In \emph{{14th Marcel Grossmann Meeting on Recent Developments in
  Theoretical and Experimental General Relativity, Astrophysics, and
  Relativistic Field Theories}}, 1 2016.
\newblock \doi{10.1142/9789813226609_0100}.

\bibitem[Hou et~al.(2018)Hou, Gong, and Liu]{Hou:2017bqj}
Shaoqi Hou, Yungui Gong, and Yunqi Liu.
\newblock {Polarizations of Gravitational Waves in Horndeski Theory}.
\newblock \emph{Eur. Phys. J. C}, 78\penalty0 (5):\penalty0 378, 2018.
\newblock \doi{10.1140/epjc/s10052-018-5869-y}.

\bibitem[Eardley et~al.(1973)Eardley, Lee, and Lightman]{Eardley:1974nw}
D.~M. Eardley, D.~L. Lee, and A.~P. Lightman.
\newblock {Gravitational-wave observations as a tool for testing relativistic
  gravity}.
\newblock \emph{Phys. Rev. D}, 8:\penalty0 3308--3321, 1973.
\newblock \doi{10.1103/PhysRevD.8.3308}.

\bibitem[Belgacem et~al.(2019)]{LISACosmologyWorkingGroup:2019mwx}
Enis Belgacem et~al.
\newblock {Testing modified gravity at cosmological distances with LISA
  standard sirens}.
\newblock \emph{JCAP}, 07:\penalty0 024, 2019.
\newblock \doi{10.1088/1475-7516/2019/07/024}.

\bibitem[Belgacem et~al.(2018)Belgacem, Dirian, Foffa, and
  Maggiore]{Belgacem:2018lbp}
Enis Belgacem, Yves Dirian, Stefano Foffa, and Michele Maggiore.
\newblock {Modified gravitational-wave propagation and standard sirens}.
\newblock \emph{Phys. Rev. D}, 98\penalty0 (2):\penalty0 023510, 2018.
\newblock \doi{10.1103/PhysRevD.98.023510}.

\bibitem[Tasinato et~al.(2021)Tasinato, Garoffolo, Bertacca, and
  Matarrese]{Tasinato:2021wol}
Gianmassimo Tasinato, Alice Garoffolo, Daniele Bertacca, and Sabino Matarrese.
\newblock {Gravitational-wave cosmological distances in scalar-tensor theories
  of gravity}.
\newblock \emph{JCAP}, 06:\penalty0 050, 2021.
\newblock \doi{10.1088/1475-7516/2021/06/050}.

\bibitem[Mirshekari et~al.(2012)Mirshekari, Yunes, and Will]{Mirshekari:2011yq}
Saeed Mirshekari, Nicolas Yunes, and Clifford~M. Will.
\newblock {Constraining Generic Lorentz Violation and the Speed of the Graviton
  with Gravitational Waves}.
\newblock \emph{Phys. Rev. D}, 85:\penalty0 024041, 2012.
\newblock \doi{10.1103/PhysRevD.85.024041}.

\bibitem[Punturo et~al.(2010)]{Punturo:2010zz}
M.~Punturo et~al.
\newblock {The Einstein Telescope: A third-generation gravitational wave
  observatory}.
\newblock \emph{Class. Quant. Grav.}, 27:\penalty0 194002, 2010.
\newblock \doi{10.1088/0264-9381/27/19/194002}.

\bibitem[Reitze et~al.(2019)]{Reitze:2019iox}
David Reitze et~al.
\newblock {Cosmic Explorer: The U.S. Contribution to Gravitational-Wave
  Astronomy beyond LIGO}.
\newblock \emph{Bull. Am. Astron. Soc.}, 51\penalty0 (7):\penalty0 035, 2019.

\bibitem[Abbott et~al.(2017)]{LIGOScientific:2017vwq}
B.~P. Abbott et~al.
\newblock {GW170817: Observation of Gravitational Waves from a Binary Neutron
  Star Inspiral}.
\newblock \emph{Phys. Rev. Lett.}, 119\penalty0 (16):\penalty0 161101, 2017.
\newblock \doi{10.1103/PhysRevLett.119.161101}.

\bibitem[de~Rham and Melville(2018)]{deRham:2018red}
Claudia de~Rham and Scott Melville.
\newblock {Gravitational Rainbows: LIGO and Dark Energy at its Cutoff}.
\newblock \emph{Phys. Rev. Lett.}, 121\penalty0 (22):\penalty0 221101, 2018.
\newblock \doi{10.1103/PhysRevLett.121.221101}.

\bibitem[Adelberger et~al.(2003)Adelberger, Heckel, and
  Nelson]{Adelberger:2003zx}
E.~G. Adelberger, Blayne~R. Heckel, and A.~E. Nelson.
\newblock {Tests of the gravitational inverse square law}.
\newblock \emph{Ann. Rev. Nucl. Part. Sci.}, 53:\penalty0 77--121, 2003.
\newblock \doi{10.1146/annurev.nucl.53.041002.110503}.

\bibitem[Will(2014)]{Will:2014kxa}
Clifford~M. Will.
\newblock {The Confrontation between General Relativity and Experiment}.
\newblock \emph{Living Rev. Rel.}, 17:\penalty0 4, 2014.
\newblock \doi{10.12942/lrr-2014-4}.

\bibitem[Jain and Khoury(2010)]{Jain:2010ka}
Bhuvnesh Jain and Justin Khoury.
\newblock {Cosmological Tests of Gravity}.
\newblock \emph{Annals Phys.}, 325:\penalty0 1479--1516, 2010.
\newblock \doi{10.1016/j.aop.2010.04.002}.

\bibitem[Koyama(2016)]{Koyama:2015vza}
Kazuya Koyama.
\newblock {Cosmological Tests of Modified Gravity}.
\newblock \emph{Rept. Prog. Phys.}, 79\penalty0 (4):\penalty0 046902, 2016.
\newblock \doi{10.1088/0034-4885/79/4/046902}.

\bibitem[Baker et~al.(2021)]{Baker:2019gxo}
Tessa Baker et~al.
\newblock {Novel Probes Project: Tests of gravity on astrophysical scales}.
\newblock \emph{Rev. Mod. Phys.}, 93\penalty0 (1):\penalty0 015003, 2021.
\newblock \doi{10.1103/RevModPhys.93.015003}.

\bibitem[Ezquiaga and Zumalac\'arregui(2020)]{Ezquiaga:2020dao}
Jose~Mar\'\i{}a Ezquiaga and Miguel Zumalac\'arregui.
\newblock {Gravitational wave lensing beyond general relativity: birefringence,
  echoes and shadows}.
\newblock \emph{Phys. Rev. D}, 102\penalty0 (12):\penalty0 124048, 2020.
\newblock \doi{10.1103/PhysRevD.102.124048}.

\bibitem[Brans and Dicke(1961)]{Brans:1961sx}
C.~Brans and R.~H. Dicke.
\newblock {Mach's principle and a relativistic theory of gravitation}.
\newblock \emph{Phys. Rev.}, 124:\penalty0 925--935, 1961.
\newblock \doi{10.1103/PhysRev.124.925}.

\bibitem[De~Felice and Tsujikawa(2010)]{DeFelice:2010jn}
Antonio De~Felice and Shinji Tsujikawa.
\newblock {Generalized Brans-Dicke theories}.
\newblock \emph{JCAP}, 07:\penalty0 024, 2010.
\newblock \doi{10.1088/1475-7516/2010/07/024}.

\bibitem[Vainshtein(1972)]{Vainshtein:1972sx}
A.~I. Vainshtein.
\newblock {To the problem of nonvanishing gravitation mass}.
\newblock \emph{Phys. Lett. B}, 39:\penalty0 393--394, 1972.
\newblock \doi{10.1016/0370-2693(72)90147-5}.

\bibitem[Babichev and Deffayet(2013)]{Babichev:2013usa}
Eugeny Babichev and C\'edric Deffayet.
\newblock {An introduction to the Vainshtein mechanism}.
\newblock \emph{Class. Quant. Grav.}, 30:\penalty0 184001, 2013.
\newblock \doi{10.1088/0264-9381/30/18/184001}.

\bibitem[Bardeen(1980)]{Bardeen:1980kt}
James~M. Bardeen.
\newblock {Gauge Invariant Cosmological Perturbations}.
\newblock \emph{Phys. Rev. D}, 22:\penalty0 1882--1905, 1980.
\newblock \doi{10.1103/PhysRevD.22.1882}.

\bibitem[Katsuragawa et~al.(2019)Katsuragawa, Nakamura, Ikeda, and
  Capozziello]{Katsuragawa:2019uto}
Taishi Katsuragawa, Tomohiro Nakamura, Taishi Ikeda, and Salvatore Capozziello.
\newblock {Gravitational Waves in $F(R)$ Gravity: Scalar Waves and the
  Chameleon Mechanism}.
\newblock \emph{Phys. Rev. D}, 99\penalty0 (12):\penalty0 124050, 2019.
\newblock \doi{10.1103/PhysRevD.99.124050}.

\bibitem[Khoury and Weltman(2004)]{Khoury:2003rn}
Justin Khoury and Amanda Weltman.
\newblock {Chameleon cosmology}.
\newblock \emph{Phys. Rev. D}, 69:\penalty0 044026, 2004.
\newblock \doi{10.1103/PhysRevD.69.044026}.

\bibitem[Burrage and Sakstein(2018)]{Burrage:2017qrf}
Clare Burrage and Jeremy Sakstein.
\newblock {Tests of Chameleon Gravity}.
\newblock \emph{Living Rev. Rel.}, 21\penalty0 (1):\penalty0 1, 2018.
\newblock \doi{10.1007/s41114-018-0011-x}.

\bibitem[Hinterbichler et~al.(2011)Hinterbichler, Khoury, Levy, and
  Matas]{Hinterbichler:2011ca}
Kurt Hinterbichler, Justin Khoury, Aaron Levy, and Andrew Matas.
\newblock {Symmetron Cosmology}.
\newblock \emph{Phys. Rev. D}, 84:\penalty0 103521, 2011.
\newblock \doi{10.1103/PhysRevD.84.103521}.

\bibitem[Garoffolo et~al.(2021)Garoffolo, Raveri, Silvestri, Tasinato, Carbone,
  Bertacca, and Matarrese]{Garoffolo:2020vtd}
Alice Garoffolo, Marco Raveri, Alessandra Silvestri, Gianmassimo Tasinato,
  Carmelita Carbone, Daniele Bertacca, and Sabino Matarrese.
\newblock {Detecting Dark Energy Fluctuations with Gravitational Waves}.
\newblock \emph{Phys. Rev. D}, 103\penalty0 (8):\penalty0 083506, 2021.
\newblock \doi{10.1103/PhysRevD.103.083506}.

\end{thebibliography}

\end{document}